%% file: acta-astronautica.tex
\documentclass[preprint,11pt]{elsarticle}
\usepackage[a4paper,textwidth=174mm,textheight=247mm,centering]{geometry}
\usepackage[T1]{fontenc}
\usepackage{amsmath,amssymb,booktabs,graphicx,siunitx}
\usepackage{newtxtext,newtxmath}
\usepackage{microtype,placeins,caption}
\usepackage{siunitx}
\usepackage[colorlinks=true,allcolors=blue]{hyperref}
\graphicspath{{paper/figures/}{figures/}}
\journal{Acta Astronautica}

\begin{document}
\begin{frontmatter}
\title{Detecting space objects with Earth-observation radar satellites}
\author[uit]{Juha Vierinen\corref{cor}}
\ead{juha-pekka.vierinen@uit.no}
\author[narvik]{Laura-Kristin Scholtz}
\author[narvik]{Ulrik Noer}
\author[narvik]{Theresa Rexer}
\author[irf]{Daniel Kastinen}
\author[narvik]{Pål Gunnar Ellingsen}
\cortext[cor]{Corresponding author.}
\affiliation[uit]{organization={Department of Physics and Technology, UiT The Arctic University of Norway},city={Troms\o},country={Norway}}
\affiliation[narvik]{organization={Department of Electrical Engineering, UiT The Arctic University of Norway},city={Narvik},country={Norway}}
\affiliation[irf]{organization={Swedish Institute of Space Physics},city={Kiruna},country={Sweden}}
\begin{abstract}
Millimetre-to-centimetre orbital debris poses a hazard to spacecraft, but its abundance remains difficult to measure. Spaceborne radars offer a substantial sensitivity advantage at close range: received echo power decreases with the fourth power of distance, so an otherwise identical object ten times closer produces a 10,000 times stronger echo. We investigate this opportunity using openly available Sentinel-1 radar voltages recorded on six calendar days. Catalogue-assisted processing identifies a manually validated sample of 76 objects at ranges of 41.7--492.9 km, including satellites, rocket bodies and debris fragments. A blind search recovers two selected debris transits without using their orbital predictions during detection. A short debris encounter also demonstrates how precise range and Doppler measurements can refine a catalogue orbit. Ground scatter increases the effective background and is the principal sensitivity limitation in Earth-looking observations. An ideal receiver-noise sensitivity comparison predicts detectable conducting-sphere diameters of approximately 3.4 mm for Sentinel-1 and 0.6 mm for EarthCARE at 10 km range, assuming coherent integration limited by beam transit time and a 10 ms cap. These results establish the feasibility of space-object detection using existing Earth-observation radars and motivate statistical debris surveys and targeted orbit refinement following collision warnings. Observations directed away from Earth and receive windows adapted to nearby objects could improve sensitivity to small debris.
\end{abstract}
\end{frontmatter}
\section{Introduction}
\begin{figure}[tbp]
\centering\input{paper/figures/sentinel-debris.include.tex}
\caption{A space object crosses Sentinel-1's Earth-looking radar beam, providing measurements of range $R$, radial velocity $\dot R$ and beam-passage time $t_b$. The line-of-sight vector $\mathbf r=R\hat{\mathbf r}$ points from radar to object; $\mathbf v_s$ and $\mathbf v_o$ are their instantaneous velocities. Relative velocity $\mathbf u=\mathbf v_o-\mathbf v_s$ gives $\dot R=\hat{\mathbf r}\cdot\mathbf u$. The long C-band SAR axis follows the flight direction and the observing face points Earthward and to the right of flight. The spacecraft outline follows the Thales Alenia Space Italia engineering view reproduced by the European Space Agency (ESA)~\cite{mission}, with body axes from~\cite{sentinelAxes}. The beam shows one instantaneous look. Object sizes, orbital altitudes and beamwidth are exaggerated for clarity.}
\label{fig:geometry}
\end{figure}
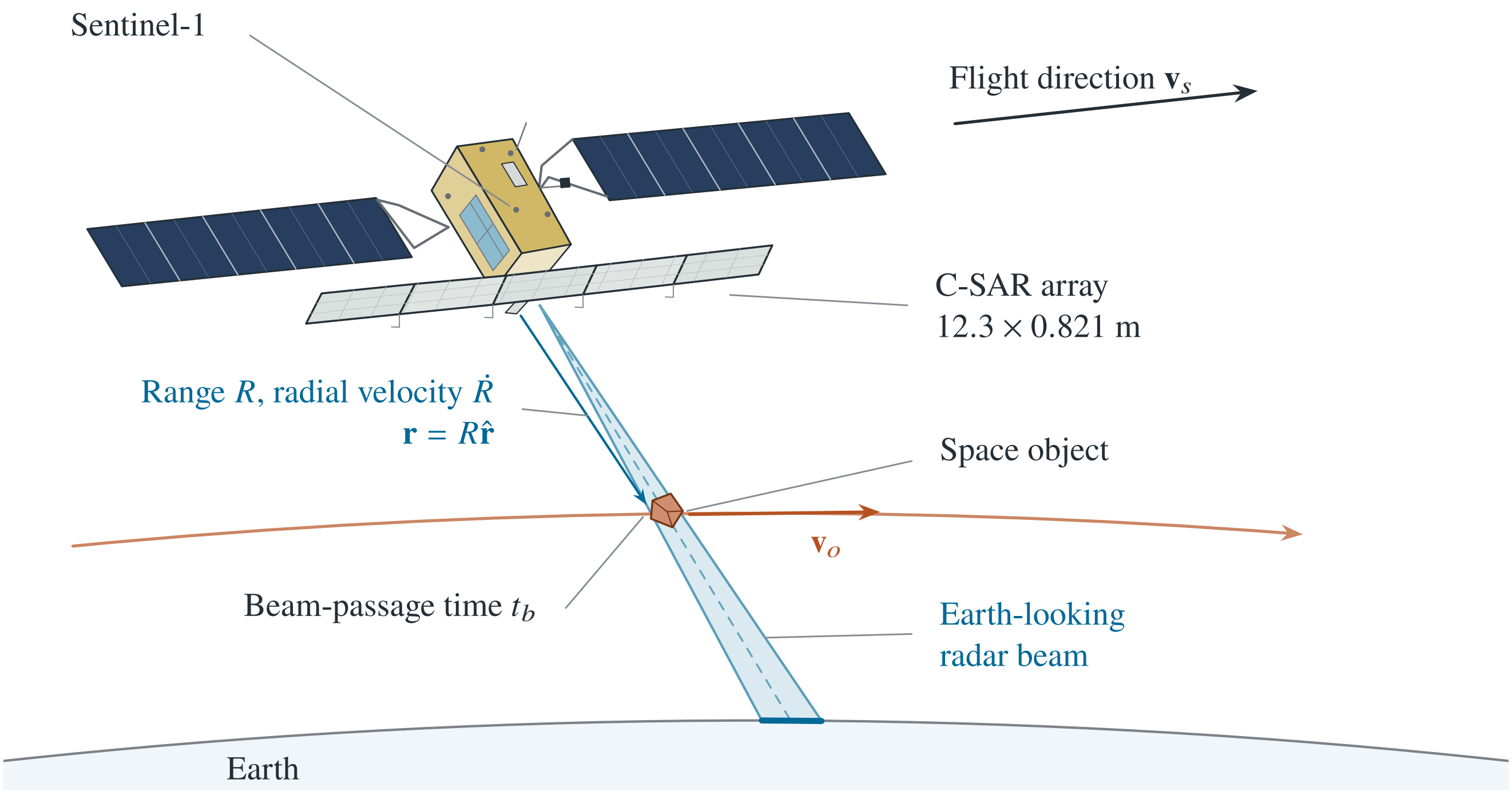

Orbital debris poses a hazard to all spacecraft in Earth orbit. Millimetre-to-centimetre fragments pose a significant impact hazard, while their small size makes routine individual tracking difficult~\cite{kennedySubcm,murrayHusir2019}. Sentinel-1A illustrates this exposure: its solar array was struck on 23 August 2016, causing impulsive orbit and attitude changes, and a permanent partial power loss. The investigation identified a man-made fragment of approximately 1 cm characteristic length as the most likely impactor~\cite{kragImpact}. Better constraints on small-fragment fluxes, sizes and encounter velocities are needed to quantify impact risk and guide spacecraft protection~\cite{ordemProcess}.
Statistical observations provide the main constraints on the subcatalogue population. Haystack and Goldstone radar measurements sample debris down to characteristic sizes of approximately 5 and 3 mm, respectively, depending on altitude and sensor configuration~\cite{kennedySubcm}. Haystack Ultrawideband Satellite Imaging Radar (HUSIR) observations resolve changes in debris flux with altitude, inclination and time~\cite{murrayHusir2019}. Together with in-situ impact measurements, these data constrain engineering environment models and spacecraft risk assessments~\cite{ordemProcess}.  Radar beam parks also characterize fragment clouds from satellite collisions and anti-satellite tests. Examples include the 2009 Iridium~33--Cosmos~2251 collision, the first accidental collision between two intact satellites~\cite{vierinenIridium2009}, and the Kosmos-1408 anti-satellite test~\cite{kosmos}. Additional viewing geometries and sensitivities would improve constraints on the small-debris environment.
An in-orbit radar can encounter debris much closer than a terrestrial sensor, providing a complementary observing geometry. Dedicated spaceborne debris radars have been studied at several scales. Cerutti-Maori et~al.\ proposed a compact instrument for detecting millimetre-size fragments at distances up to approximately 500 m~\cite{cerutti}. Ram\'{i}rez-Torres et~al.\ described a 94 GHz radar prototype and its enabling hardware~\cite{ramirez}, while Maffei et~al.\ developed a Ka-band electronically scanned pulse-Doppler architecture with range/range-rate search, pulse compression, detection and tracking~\cite{maffei}. Our QBDebris study proposed CubeSats carrying millimetre-wavelength frequency-modulated continuous-wave (FMCW) radars, with overlapping observations to improve debris trajectory estimates~\cite{qbdebris, Retselis2026}. Satellite-borne radar has also been proposed for collision avoidance~\cite{katz}. Anttonen et~al.\ investigated debris sensing through intersatellite communication signals, including high-velocity estimation and constellation cooperation~\cite{anttonen}.
Existing Earth-observation radars offer an opportunity to test these ideas with recorded measurements. Sentinel-1, the same satellite type impacted in 2016, operates a C-band synthetic aperture radar (SAR) at 5.405 GHz with approximately 4 kW peak transmitted power~\cite{torres,mission}. Its openly available Level-0 products preserve received waveforms~\cite{level0}. Standard SAR processing focuses stationary ground targets. Reprocessing the voltages with an orbital motion model focuses a space-object passage through the Earth-looking beam (Fig.~\ref{fig:geometry}).
Here we demonstrate space-object detection in Sentinel-1 measurements using catalogue-assisted matched filtering and a blind range-migration search. We present manually validated satellite and debris returns, compare blind and assisted recovery on the same recorded pulses, and estimate radar cross section (RCS) from fitted beam passages. A receiver-noise sensitivity study then examines the potential for statistical surveys with existing orbital radar instruments.
\section{Data}
\subsection{Recorded voltages and acquisition modes}
We screened catalogue conjunctions over 11 calendar days, 26 October – 5 November 2026. The waveform analysis uses approximately 1\,271 s (21.2 min) of Sentinel-1C and Sentinel-1D raw voltages recorded on six calendar days, 29 October – 3 November. This duration counts overlapping excerpts and repeated searches only once. The Level-0 products contain the original complex echo samples and the packet metadata needed to reconstruct the recordings~\cite{level0,cdseLevel0}.
The receiver digitizes the signal at 300 MHz, then digitally filters and downsamples it before recording~\cite{mission}. The recorded complex sampling rate depends on the acquisition settings. For the DELTA 1 DEB IW2 and FENGYUN 1C DEB EW4 examples, it is 54.596 and 17.324 million complex samples per second, respectively, corresponding to range-sample spacings of 2.746 and 8.653 m. We use the recorded sampling rates when reconstructing the waveform. In both catalogue-assisted and blind processing, eight-tap Lanczos windowed-sinc interpolation evaluates the complex pulse-compressed signal at fractional delays between these recorded samples. This keeps the moving echo aligned across pulses during coherent integration.
Before downlink, Flexible Dynamic Block Adaptive Quantization (FDBAQ) reduces the data volume by quantizing the in-phase and quadrature components, adapting the bit allocation to signal power, and applying Huffman coding~\cite{fdbaqSpec}. Both illustrated debris acquisitions use nominal FDBAQ mode 0 with 256 complex samples per block. We reconstruct the complex samples using the recorded block thresholds, bit-rate codes and specified reconstruction tables. Quantization rounds amplitudes to a finite set of levels, introducing reconstruction errors in the voltages; the subsequent Huffman coding is lossless.
The metadata also give reception times, intervals between transmitted pulses, receive-window start times and transmitted chirp settings. The chirp settings describe the frequency sweep of each transmitted pulse and define the linear-frequency-modulated (LFM) waveform used for matching. We process echo packets separately from noise and calibration packets. All times are given in Coordinated Universal Time (UTC). Polarization labels specify the transmitted orientation first and the received orientation second: HH and VV are horizontal and vertical co-polarization, while HV and VH are the cross-polarized channels.
The acquisition mode determines how the antenna illuminates the scene. Stripmap (SM) observes one ground strip with nominally fixed azimuth and elevation pointing, whereas Interferometric Wide Swath (IW) and Extra Wide Swath (EW) cycle through three and five elevation subswaths, respectively. Both use Terrain Observation by Progressive Scans (TOPS), in which the beam scans electronically from backward-looking to forward-looking during each burst, then resets for the next subswath. This steering distributes illumination more uniformly across ground targets than conventional scanning SAR (ScanSAR), reducing along-track brightness variations. ESA gives azimuth steering limits of $\pm0.6^\circ$ for IW and $\pm0.8^\circ$ for EW, with characteristic burst durations of approximately 0.82 and 0.54 s, respectively~\cite{mission,tops}. These correspond to approximate sweep rates of $1.5^\circ$\,s$^{-1}$ and $3^\circ$\,s$^{-1}$.
A space object is illuminated only while it lies within the active beam. For example, the EW2 bursts used here last approximately 0.546 s and recur about every 3 s as the radar cycles through the subswaths. Within a burst, the object's motion relative to the scanning beam determines its passage duration. The burst end, departure from the elevation beam and the receive-window timing determine which echo pulses are recorded.
Sentinel-1's high-rate radar operating allowance is 30 minutes per 98.6-minute orbit~\cite{s1operations}. If fully used, this provides approximately 7.3 hours of radar operation per day per satellite in SM, IW and EW modes, or 14.6 satellite-hours per day for two satellites. Wave-mode observations add radar activity over open oceans. These operating periods define the opportunities to examine catalogue-predicted conjunctions in routine Earth-observation recordings.
\subsection{Pulse timing and recorded range}
The pulse repetition interval $T_r$ is the time between successive transmitted pulses within a burst and Table~\ref{tab:gates} summarizes the representative recorded settings. Additionally, the instrument supports pulse durations of \SIrange{5}{100}{\us}~\cite{mission}. The transmitted pulse duration and receiver recording times determine how much of a nearby object's echo is captured.
\begin{table}[htbp]\centering
\caption{Representative pulse timing, range ambiguity spacing and minimum ranges for recording \SI{10}{\percent} or all of an echo pulse.}
\label{tab:gates}
\begin{tabular}{@{}lrrrrr@{}}\toprule
Mode & Pulse [$\mu$s] & $T_r$ [$\mu$s] & $c\,T_r/2$ [km] & 10\% of echo [km] & Full echo [km]\\\midrule
EW1 & 30.29 & 606.82 & 90.96 & 6.15 & 10.24\\
EW2 & 25.84 & 515.66 & 77.29 & 10.76 & 14.24\\
EW3 & 30.35 & 606.88 & 90.97 & 4.40 & 8.49\\
EW4 & 26.35 & 526.90 & 78.98 & 4.70 & 8.26\\
EW5 & 30.66 & 613.25 & 91.92 & 3.91 & 8.04\\
IW1 & 52.40 & 582.37 & 87.29 & 6.24 & 13.31\\
IW2 & 62.00 & 688.88 & 103.26 & 10.98 & 19.34\\
IW3 & 53.39 & 593.18 & 88.92 & 4.69 & 11.89\\
SM/S3 & 44.17 & 519.49 & 77.87 & 8.58 & 14.54\\\bottomrule
\end{tabular}
\end{table}
In the sampled recordings, a complete echo pulse can be captured at ranges as short as 8.04 km in EW and 11.89 km in IW. At shorter ranges, only part of the echo falls within the recording window. The minimum ranges for capturing \SI{10}{\percent} of the pulse duration are 3.91 km in EW and 4.69 km in IW. For a constant-amplitude pulse, this also corresponds to \SI{10}{\percent} of its energy.
Repeated transmission of similar pulses can make it unclear which pulse produced a given echo. Choosing a pulse transmitted one interval earlier increases the inferred round-trip travel time by $T_r$, and therefore the range by $cT_r/2$, where $c$ is the speed of light. For the pulse timings in Table~\ref{tab:gates} the possible ranges are separated by approximately \SIrange{77}{103}{\km}.
\subsection{Orbit predictions and sample selection}
The target positions are predicted using daily U.S. Space Force general-perturbation (GP) orbital elements, distributed as two-line elements (TLEs). We propagate the target orbits with Simplified General Perturbations 4 (SGP4)~\cite{vallado}. Combining these predictions with Sentinel-1's recorded orbit and antenna pointing gives the encounter geometry during radar operation. Object identifiers follow the North American Aerospace Defense Command (NORAD) catalogue.  As the catalogue elements have no reported uncertainties, we allow, during candidate selection, the target's predicted position to shift by up to $\pm2$ s along its orbit to accommodate possible along-track errors.
Predicted conjunctions guide the selection of recordings and the catalogue-assisted search. We inspect the resulting time--range and time--Doppler maps and manually select the accepted echoes. For the RCS comparison, we use one encounter per catalogue object. If an object has several manually accepted detections, we select the encounter with the strongest echo. This sample demonstrates detection and provides examples for detailed echo fitting. Measuring detection efficiency requires a separate study with a defined sampling procedure and recovery tests.
We compare catalogue-assisted and blind processing for two debris encounters: DELTA 1 DEB (NORAD 8951), observed on 1 November 2026, and FENGYUN 1C DEB, observed on 3 November. Both methods use identical recorded voltages for each encounter.
\section{Methods}
\subsection{Range, Doppler and waveform matching}
We detect space-object echoes by coherently combining the received pulses. For each trial object trajectory, we calculate the expected echo waveform and compare it with the recorded complex voltages. When the trial follows the object's motion, it compensates for changes in the echo's arrival time and phase, allowing the signal from successive pulses to add coherently.
The expected waveform depends on the changing radar--object range. In the blind search, we describe this range using three parameters: range $R_0$, radial velocity $v$ and radial acceleration $a$, evaluated at the centre of the integration window, $t_0$. With $h=t-t_0$,
\begin{equation}
R_{\rm blind}(t_0+h)\simeq R_0+vh+\tfrac12ah^2.
\label{eq:motion}
\end{equation}
For catalogue-assisted processing, we calculate the predicted range $R_c(t)$ from the catalogue orbit and Sentinel-1's recorded orbit. We then search for corrections $\delta R_0$, $\delta v$ and $\delta a$ to the predicted range, radial velocity and radial acceleration:
\begin{equation}
R_{\rm assisted}(t_0+h)=R_c(t_0+h)
+\delta R_0+\delta v h+\tfrac12\delta a h^2.
\label{eq:range-correction}
\end{equation}
Each trial range history determines both the echo's arrival time and its phase. A range correction $\delta R(h)$ changes the round-trip delay by $2\delta R(h)/c$ and the carrier phase by $-4\pi\delta R(h)/\lambda$, where $c$ is the speed of light and $\lambda$ is the radar wavelength. The corresponding Doppler correction is $-2(\delta v+\delta a h)/\lambda$. Matching these changes keeps the echo aligned in range and phase as the object moves during coherent integration.
For complex samples $z_{pk}$, indexed by pulse $p$ and sample $k$, and a trial echo replica $q_{pk}(\boldsymbol\theta)$, we maximize the coherent matched-filter power
\begin{equation}
P(\boldsymbol\theta)=
\frac{\left|\sum_{p,k}z_{pk}q_{pk}^{*}(\boldsymbol\theta)\right|^2}
{\sum_{p,k}|q_{pk}(\boldsymbol\theta)|^2},
\label{eq:mf}
\end{equation}
where $\boldsymbol\theta$ contains the trial range, velocity and acceleration, or their corrections to the catalogue prediction. The replica uses the recorded chirp parameters and receive timing, including fractional delay and Doppler time scaling.
The relative position, velocity and acceleration, for the object (subscript $o$) and Sentinel-1 (subscript $s$) in a common inertial frame are
\begin{align}
\mathbf r&=\mathbf r_o-\mathbf r_s=R\hat{\mathbf r}\\
\mathbf u&=\mathbf v_o-\mathbf v_s\\
\mathbf b&=\mathbf a_o-\mathbf a_s
\end{align}
These give
\begin{equation}
\dot R=\hat{\mathbf r}\cdot\mathbf u,\qquad
\ddot R=\frac{|\mathbf u|^2-\dot R^2}{R}+\hat{\mathbf r}\cdot\mathbf b,
\label{eq:accel}
\end{equation}
where the first term in $\ddot R$ describes the effect of motion across the line of sight. Near closest approach, an object may move rapidly across the radar's view while its distance changes only slowly. As it passes, the viewing direction rotates rapidly and the rate of change of distance can increase sharply. The second term is the relative acceleration of the object and spacecraft projected along the line of sight. Thus a close encounter can have a small instantaneous Doppler shift but a rapidly changing Doppler frequency
\begin{equation}
f_D=-\frac{2\dot R}{\lambda},\qquad
\dot f_D=-\frac{2\ddot R}{\lambda},
\end{equation}
where positive $\dot R$ denotes an increasing distance and $\lambda=5.55$ cm is the radar wavelength.
For example, motion across the line of sight at 16 km/s contributes \SI{2560}{\m\per\s\squared} to $\ddot R$ at 100 km range, increasing to \SI{25\,600}{\m\per\s\squared} at 10 km. The same motion crosses a fixed 0.004 rad beam in approximately 25 ms at 100 km and 2.5 ms at 10 km. Close encounters therefore require short integration windows and accurate compensation for the changing range.
Our matched filter jointly fits range, radial velocity and acceleration, including the effect of Doppler on the chirped waveform. Small changes in delay and Doppler can partly compensate for one another, allowing different parameter combinations to produce similar responses.
Repeated transmission introduces two further ambiguities. Doppler aliasing arises because frequencies separated by $1/T_r$ produce the same pulse-to-pulse phase change. Range ambiguity arises when the transmitted pulse responsible for an echo is uncertain: assigning the echo to a pulse transmitted one interval earlier increases its inferred range by $cT_r/2$, where $c$ is the speed of light. We compare the possible solutions using the recorded waveforms and, for catalogue-assisted processing, the predicted orbit.
The physical range and Doppler are related to the apparent pulse-compressed coordinates by
\begin{equation}
f_D=f_{\rm wrapped}+n\,f_r,\qquad
R\simeq R_{\rm app}+\frac{cf_D}{2K}+m\frac{c\,T_r}{2},
\label{eq:aliases}
\end{equation}
where $f_r=1/T_r$ is the pulse repetition frequency, $T_r$ is the interval between transmitted pulses, $f_{\rm wrapped}$ is the Doppler frequency measured within one interval of width $f_r$, and $R_{\rm app}$ is the apparent range after pulse compression. The integer $n$ selects a possible Doppler frequency, and $m$ selects the transmitted pulse responsible for the echo. Ranges are in metres and frequencies in hertz. The pulse interval $T_r$ is in seconds, the signed chirp slope $K$ in Hz\,s$^{-1}$, and $c$ in m\,s$^{-1}$. The recorded pulse sequence and receive times constrain the allowed combinations of $m$ and $n$, which we retain when interpreting candidate trajectories.
We also report the coherent fraction,
\begin{equation}
z_p = \sum_{k}z_{pk}q_{pk}^{*}(\hat{\boldsymbol\theta}),\qquad
C=\frac{|\sum_p z_p|^2}{N_p\sum_p|z_p|^2},
\end{equation}
Here $\hat{\boldsymbol\theta}$ is the common trajectory estimate that maximizes the coherent power in Eq.~\ref{eq:mf}. The complex matched-filter output $z_p$ combines the recorded samples $z_{pk}$ for pulse $p$ with the waveform replica evaluated at this estimate, and $N_p$ is the number of pulses. The normalization compares the coherent power with the maximum possible for the same total pulse power. Values close to one indicate constructive addition and efficient coherent integration; values close to zero indicate cancellation between pulse outputs.
\subsection{Searches and detection}
The catalogue-assisted search tests small corrections around the predicted object trajectory. Without that prior, the blind search tests the recorded range interval, range rates within \SI{\pm16}{\km\per\s} and accelerations within \SI{\pm50\,000}{\m\per\s\squared}. Both use 16-pulse coherent processing intervals (CPIs). The starting ranges remain on the recorded sample grid. For the blind search, we subdivide the migration-rate grid eightfold relative to a one-range-sample displacement over the CPI, giving displacement increments of approximately 0.34 m for DELTA 1 DEB and 1.05 m for FENGYUN 1C DEB. Fourfold zero-padding of the 16-pulse sequence gives a 64-bin fast Fourier transform (FFT), sampling the wrapped Doppler response more densely. Starting from each candidate, we adjust its range, radial velocity and acceleration to maximize the coherent matched-filter power in the original complex voltages. The computational cost limits this blind-search demonstration to the two selected debris encounters.
For the \emph{blind search}, the grid used for the DELTA 1 DEB encounter contains 23\,884 starting ranges, 977 range rates and 195 accelerations, giving $H_{\rm blind}\simeq4.6\times10^9$ combinations per 16-pulse window. With \SI{50}{\percent} window overlap, this corresponds to approximately $8.3\times10^{11}$ blind-search combinations per second of continuously recorded pulses.
For comparison, a \emph{catalogue-assisted} reference grid searches corrections within $\pm2$ km in range, \SI{\pm100}{\m\per\s} in range rate and \SI{\pm100}{\m\per\s\squared} in acceleration. At the same acquisition settings, its 1465 ranges, three range rates and three accelerations give $H_{\rm assisted}=13\,185$ combinations per window. At the same window cadence, this is approximately $2.4\times10^6$ assisted-search combinations per second, for each catalogue timing hypothesis. The blind grid therefore contains approximately $3.5\times10^5$ times more combinations than this assisted reference grid.
Each range--motion combination also tests 64 wrapped Doppler frequencies. Thus each window evaluates $64H$ filter responses, with the FFT calculating the 64 Doppler responses together. The $H$ counts describe range--motion trials; the arithmetic work after pulse compression scales as $\mathcal O\!\left[H(P+F\log_2 F)\right]$ per window, where $P=16$, $F=64$, and $H$ is $H_{\rm blind}$ or $H_{\rm assisted}$ for the respective search.
We select candidate peaks relative to the surrounding background using cell averaging, following constant false-alarm rate (CFAR) processing~\cite{skolnik2001}. The threshold is 6 dB above the local mean after maximization over motion and Doppler, chosen using the two known echoes. Nearby peaks are grouped into candidate events, with repeated appearances and alternative ambiguity branches counted together. All primary measurements use the original voltages with ground cancellation disabled.
\subsection{Orbit refinement from a short encounter}
We use an idealized DELTA 1 DEB example to test whether one short radar passage can reduce orbit uncertainty half an orbit later. The passage provides three types of information: range, Doppler and an angular constraint obtained by fitting the echo envelope to the known antenna pattern and TOPS pointing history. The envelope locates the object within the narrow azimuth beam.
We divide the full passage into 19 nonoverlapping 16-pulse windows spanning approximately 0.20 s and use all of them to fit the power envelope. Five windows also provide range and Doppler measurements: these have echo SNR above 6 dB on the fixed initial trajectory, before local waveform refinement. We hold the TOPS pointing and the $0.23^\circ$ one-way beamwidth fixed, and fit a common echo amplitude together with the orbit. The modeled power is $\widehat P=B+A g$, where $B$ is the off-echo background and $g$ is the two-way beam response averaged over each window. From the background  the power-fit weights are set. The range--Doppler likelihood retains the waveform-fit covariance and tested Doppler alternatives, accounting for chirp coupling.
We update the six SGP4 mean orbital elements by differential correction~\cite{vallado2008od}, keeping the catalogue epoch and drag terms fixed. Predicted measurements combine the target orbit with Sentinel-1's annotated orbit and account for signal travel time. A Gaussian Cartesian-state prior favours small corrections to the catalogue orbit~\cite{montaruli2024od}. Initial standard deviations of 34.3 m per position coordinate and \SI{0.0343}{\m\per\s} per velocity coordinate give a three-dimensional root-mean-square (RMS) position spread of approximately 500 m after half an orbit, or 49 min.
We compare updates using range and Doppler alone and using all three sources of information, with the same prior in both cases. Four Markov chain Monte Carlo (MCMC) chains with Metropolis updates sample the nonlinear posterior over the six mean elements~\cite{virtanen2013bayes}, retaining the Doppler alternatives. Each fit provides 32,000 retained orbit Metropolis-Hastings Markov chain Monte-Carlo  samples \cite{gelman2013bda}. We propagate every sampled orbit to the same future epoch and compare its position spread with the propagated prior. This measures the information supplied by the passage under the assumed pointing, noise and prior models.
\subsection{Beam fitting and RCS estimation}
\label{sec:beam-rcs}
The return brightens and fades as an object crosses the antenna beam. We fit this envelope to account for changing azimuth gain before estimating RCS~\cite{kosmos}.
For object and beam azimuths $\psi_o(t)$ and $\psi_b(t)$ in the antenna frame, the illumination time is approximately
\begin{equation}
T_{\rm illum}\simeq\frac{\theta_{\rm az}}{|\dot\psi_o-\dot\psi_b|},
\label{eq:tops-dwell}
\end{equation}
where $\theta_{\rm az}$ is the one-way beamwidth and the relative angular rate is assumed constant during the crossing. The TOPS scan alone would sweep a $0.23^\circ$ beam past an object with negligible angular motion in approximately \SIrange{0.08}{0.15}{\s}. An orbiting object's own angular motion can substantially change this duration.
We fit $P(t)=B+A g(t)$ in linear power. The background $B$ is estimated from up to 64 off-echo range gates at offsets of \SIrange{250}{2234}{\m}, using the same pulses and motion compensation. The response $g(t)$ is a Gaussian two-way azimuth pattern averaged coherently over each window, with one-way half-power width $0.23^\circ$. For the RCS estimates, the trajectory is fixed; amplitude and azimuth pointing offset are fitted, with a common offset for both polarizations. Total power/background is $Q=P/B$, and echo/background is $Q-1$.
The coherent radar equation relates RCS to receiver-noise signal-to-noise ratio (SNR) $\rho$:
\begin{equation}
\rho=\frac{P_{\rm pk}G_tG_r\lambda^2\sigma T_{\rm tx}}
{(4\pi)^3 L R^4 k_B T_{\rm sys}},\qquad
T_{\rm tx}=\sum_p T_p,
\label{eq:event-rcs}
\end{equation}
where $P_{\rm pk}$ is peak transmitted power, $G_t,G_r$ are gains toward the object, $T_p$ is pulse duration, $T_{\rm sys}$ is system temperature, $k_B$ is Boltzmann's constant and $L$ represents other losses. We invert this equation with $\rho=Q-1$ to estimate the RCS and adopt a 606 K thermal background, combining a 290 K antenna temperature with the published 3.2 dB receiver noise figure, and a reference peak power of 4075 W, specified for dual-polarization IW~\cite{torres,mission}. The nominal one-way gain of 43.56 dBi is calculated from the published antenna dimensions~\cite{torres} with an assumed aperture efficiency of 0.55. We assume unit elevation response and $L=1$. The fit uses the upper 6 dB of the modeled echo, correcting for beam gain, range and transmitted pulse duration. These assumptions set the absolute RCS scale; the measured background includes ground scatter.
\subsection{Receiver-noise sensitivity comparison}
\label{sec:receiver-sensitivity}
A nearby object produces a stronger echo but also crosses the antenna beam more quickly, leaving less time to collect its signal. Our sensitivity comparison accounts for both effects. We express the result as a noise-equivalent diameter $D_n$: the smallest perfectly conducting sphere giving unit output SNR under the assumed observing conditions. A smaller diameter indicates greater sensitivity. We use the full sphere-scattering series~\cite{stratton} and the radar equation (Eq.~\ref{eq:event-rcs}), with instrument parameters in Table~\ref{tab:radars}.
The comparison includes Sentinel-1, the National Aeronautics and Space Administration (NASA)--Indian Space Research Organisation (ISRO) Synthetic Aperture Radar (NISAR), TerraSAR-X, Sentinel-3's Synthetic Aperture Radar Altimeter (SRAL) and EarthCARE's Cloud Profiling Radar (CPR). We assume central crossings at transverse speed $v_\perp=10$ km/s, constant boresight gain, complete echo recording and perfect motion compensation. The coherent integration time is
\begin{equation}
T_{\rm int}=\min(10\,\mathrm{ms},R\theta_b/v_\perp),\qquad T_{\rm tx}=\delta T_{\rm int},
\label{eq:sensitivity-dwell}
\end{equation}
where $\theta_b$ is the crossing beamwidth and $\delta$ the transmit duty factor. We set $\rho=1$ and the loss factor $L=2$ (3 dB). Thus the calculation accounts for the shorter beam dwell at nearby ranges and the fraction of that dwell spent transmitting. These assumptions define the best-case comparison.
\begin{table}[htbp]\centering\small
\caption{Radar parameters and ideal unit-SNR conducting-sphere diameters at 10 and 50 km. * marks estimated power or gain. Noise temperatures, duty factors and crossing widths are adopted comparison values; sources and calculation assumptions are given in the text.}
\label{tab:radars}
\begin{tabular}{@{}lrrrrrrrr@{}}\toprule
Radar & GHz & Peak kW & Gain dBi & \shortstack{$T_{\rm sys}$\\(K)} & \shortstack{$\delta$\\(\%)} & \shortstack{$\theta_b$\\($^\circ$)} & \shortstack{$D_n$\\10 km [mm]} & \shortstack{$D_n$\\50 km [mm]}\\\midrule
Sentinel-1 & 5.405 & 4.368 & 43.6* & 606 & 9.0 & 0.26 & 3.43 & 8.87\\
NISAR L & 1.258 & 4.000* & 41.4* & 600 & 10.0 & 1.3 & 5.74 & 16.82\\
NISAR S & 3.200 & 37.857 & 49.5* & 600 & 10.0 & 0.5 & 1.59 & 4.54\\
TerraSAR-X & 9.650 & 2.260 & 43.8* & 600 & 10.0 & 0.33 & 2.92 & 8.37\\
Sentinel-3 SRAL Ku & 13.575 & 0.009 & 41.5 & 600 & 10.0 & 1.2 & 15.85 & 546.11\\
EarthCARE CPR & 94.050 & 1.430 & 65.2* & 2000 & 2.15 & 0.095 & 0.56 & 4.72\\\bottomrule
\end{tabular}
\end{table}
Sentinel-1's frequency, summed transmit/receive-module output power, $12.3\times0.821$ m aperture and IW duty factor follow Torres et~al.~\cite{torres}. The NISAR L-band power is assumed. NISAR frequencies follow the ISRO specification~\cite{nisarFormat}, with a 12 m reflector~\cite{nisarFeed}. Its S-band power follows the Space Frequency Coordination Group compilation~\cite{radarRF}. TerraSAR-X frequency and power follow the same source, and it has a $4.8\times0.7$ m aperture~\cite{tsxAntenna}. The SRAL frequency, power and gain follow the ESA mission monograph~\cite{sralSpec}.
EarthCARE's frequency, minimum specified power, 2.5 m aperture and beamwidth follow the Japan Aerospace Exploration Agency (JAXA)~\cite{earthcare2020}. Its \SI{2.15}{\percent} duty factor corresponds to 3.3 $\mu$s pulses at approximately 6.5 kHz, within the specified \SIrange{6.1}{7.5}{\kilo\Hz} range. Apart from SRAL's specified gain, we calculate $G=4\pi(0.55)A/\lambda^2$ from aperture area $A$, assuming efficiency 0.55 and the full aperture on transmit and receive.
The Sentinel-1 noise temperature combines a 290 K antenna temperature with the receiver contribution from the published 3.2 dB noise figure~\cite{torres}, giving 606 K. Other noise temperatures, duty factors and crossing widths are illustrative assumptions unless specified above.
\section{Results}
\input{paper/data/human-validated-rcs-counts.tex}
The results address three questions: which space objects are visible in the recorded data, whether their echoes can be found without an ephemeris, and how the radar measurements can improve an orbit prediction. We begin with catalogue-associated detections and their echo strengths, then examine blind recovery and orbit refinement. Finally, we compare the ideal sensitivities of several orbital radars to assess the potential for detecting smaller debris.
\subsection{Satellite and debris detections}
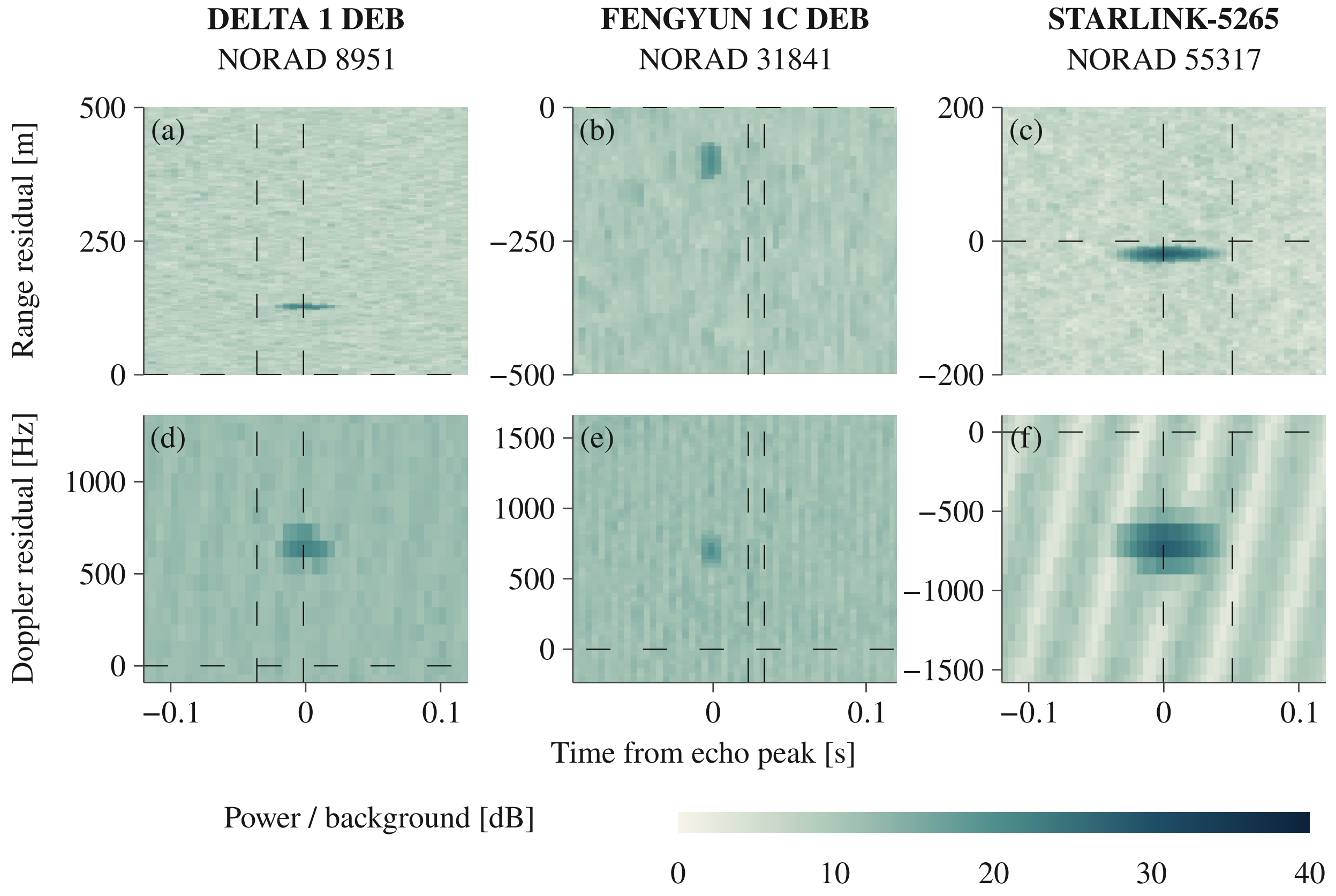
\begin{figure}[htbp]\centering
\input{paper/figures/assisted-comparison.include.tex}
\caption{Catalogue-assisted detections of two debris fragments and an operational satellite. Columns show DELTA 1 DEB, FENGYUN 1C DEB and STARLINK-5265. The upper panels show measured range minus catalogue-predicted range; the lower panels show the corresponding Doppler difference, wrapped into the pulse-sampling interval. Compact patches show the object echoes, and colours indicate power relative to local background in dB. Horizontal dashed lines mark zero residual; vertical dashed lines mark the predicted half-power beam boundaries. Time is relative to each echo peak. Each estimate integrates 16 pulses coherently.}
\label{fig:delta-assisted}
\end{figure}

Catalogue-assisted processing identifies operational satellites, rocket bodies and debris fragments in the recorded data. The manually validated sample contains \HumanRcsObjectCount{} distinct catalogue objects at fitted ranges of \SIrange{41.7}{492.9}{\km}, drawn from 1\,271 s of raw-voltage excerpts.
Figure~\ref{fig:delta-assisted} shows what these detections look like, using two nearby debris fragments and a more distant operational satellite. An echo following the catalogue prediction would lie at zero in both the range and Doppler residual panels; offsets show where the measured echo differs from that prediction. DELTA 1 DEB (NORAD 8951) at 65.7 km and FENGYUN 1C DEB (NORAD 31841) at 41.7 km appear as compact echoes. FENGYUN 1C DEB is the closest object in the validated sample and is detected in both HH and HV. STARLINK-5265, at 277.9 km, illustrates a strong satellite return, with a peak VV echo SNR of approximately 31 dB.
\FloatBarrier
\subsection{Echo strength and polarization}
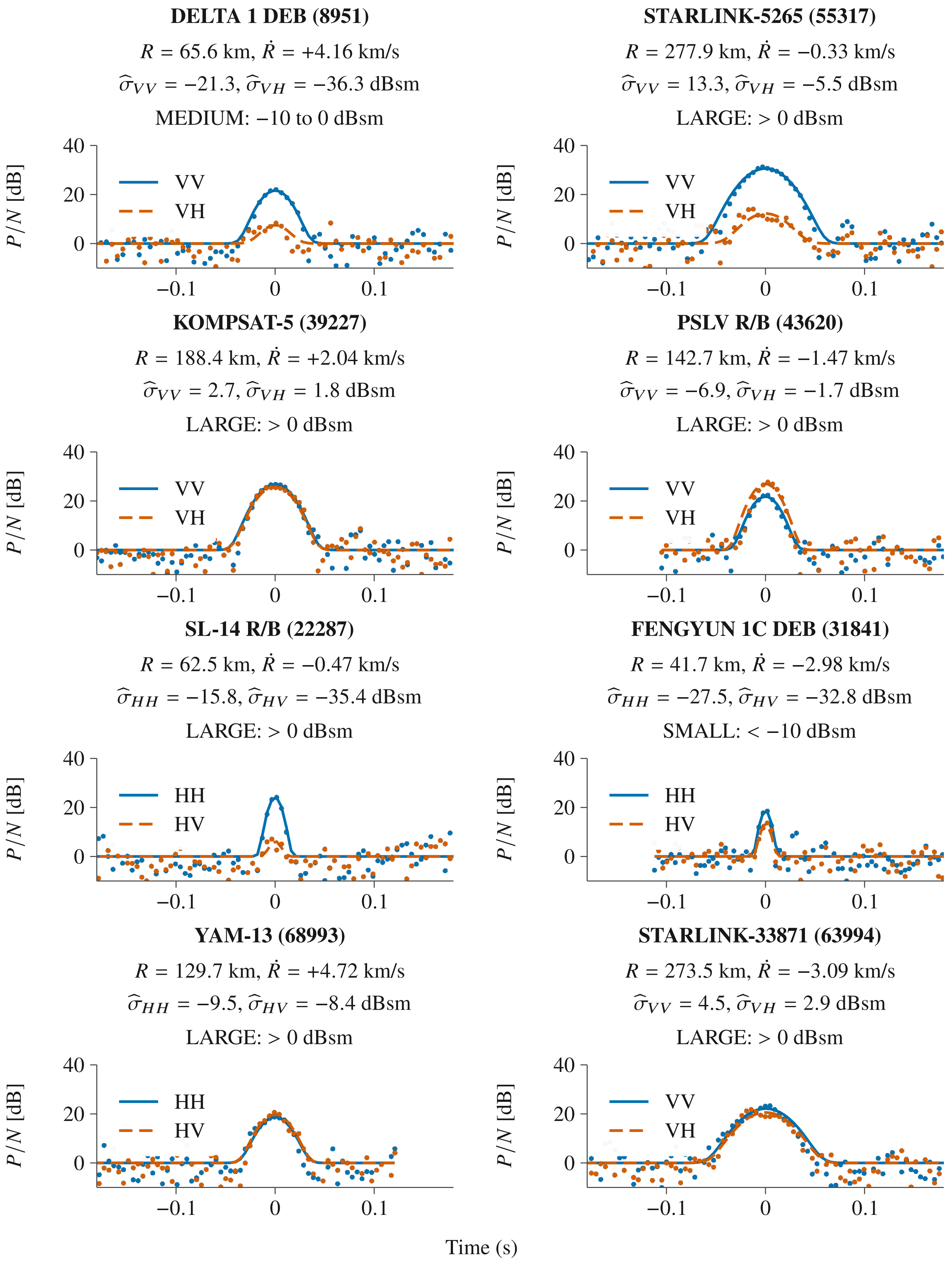
\begin{figure}[tbp]
\centering\input{paper/figures/beam-examples.include.tex}
\caption{Echo strength during eight object passages through the antenna beam. Points show measured power relative to background, using coherent integrations of 16 pulses; curves show the fitted beam-crossing envelopes. Colours distinguish the recorded polarizations. Zero dB corresponds to the background level, and time zero marks the fitted beam-centre crossing. Annotations give range $R$, radial velocity $\dot R$, estimated RCS $\hat\sigma$ for each polarization, and the historical NORAD RCS class. RCS estimates use the portion of each fitted echo within 6 dB of its peak and the calibration assumptions in Section~\ref{sec:beam-rcs}.}
\label{fig:delta-envelope}
\end{figure}

\begin{figure}[htbp]
\centering\input{paper/figures/human-validated-rcs.include.tex}
\caption{Estimated radar cross sections for the \HumanRcsObjectCount{} catalogue objects, grouped by historical NORAD RCS class and polarization. Each point represents one encounter in one polarization; horizontal offsets separate the polarization channels. The histograms at right summarize the same estimates in 5 dB bins, and legend counts give the number of estimates per channel. NORAD classes correspond to areas below 0.1 m$^2$ (SMALL), 0.1--1 m$^2$ (MEDIUM) and above 1 m$^2$ (LARGE). These classes provide a broad comparison with the RCS measured at the observing frequency, polarization and viewing geometry. Each object contributes up to two polarization-specific estimates.}
\label{fig:human-rcs}
\end{figure}
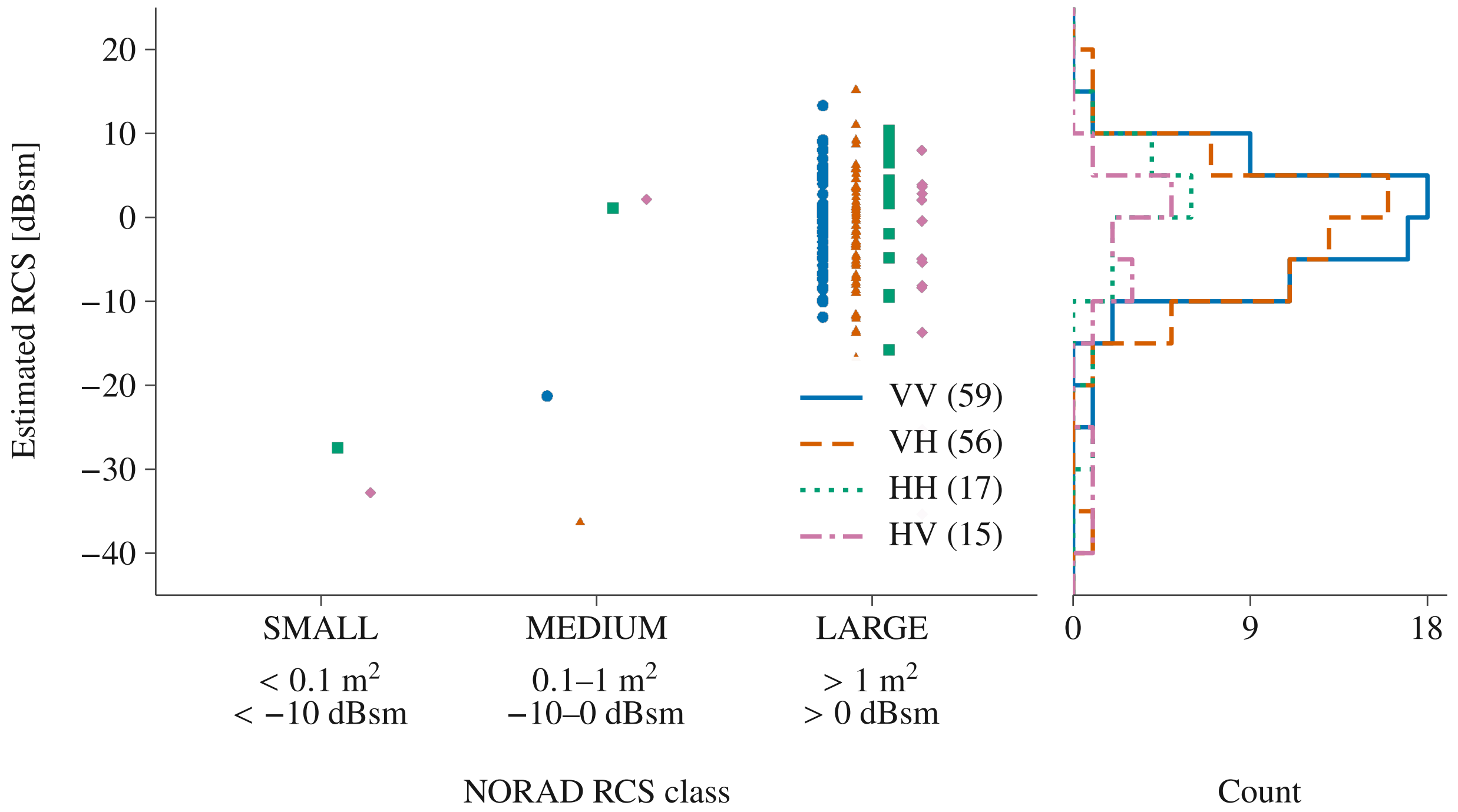

The echo brightens and fades as each object crosses the antenna beam. Figure~\ref{fig:delta-envelope} shows eight examples and the fitted envelopes used to estimate RCS. The passages differ in duration and strength, and the two recorded polarizations can give substantially different returns from the same object.
Both debris examples return more power in the transmitted polarization than in the orthogonal polarization. DELTA 1 DEB has estimated RCS values of $-21.3$ dBsm in VV and $-36.3$ dBsm in VH, a difference of 15 dB. FENGYUN 1C DEB gives $-27.5$ dBsm in HH and $-32.8$ dBsm in HV, a difference of 5.3 dB. Thus the inferred RCS depends on polarization as well as the object and viewing geometry.
Figure~\ref{fig:human-rcs} places these examples within the full sample, grouping the estimates by NORAD RCS class and polarization. The \HumanRcsMeasurementCount{} estimates span $-36.3$ to $+15.1$ dBsm: 59 VV, 56 VH, 17 HH and 15 HV. Each object contributes one selected encounter and up to two polarization channels. The NORAD classes provide a broad size-related comparison, while our estimates describe the radar response during the observed encounter.
\FloatBarrier
\subsection{Blind recovery}
\label{sec:blind-results}
\begin{figure}[tbp]
\centering\input{paper/figures/blind-comparison.include.tex}
\caption{Blind recovery of DELTA 1 DEB (left) and FENGYUN 1C DEB (right), without using their ephemerides during the search. Panels (a,b) show the strongest responses across trial motions and Doppler frequencies, using 16-pulse integrations. Panels (c,d) show the echoes after applying the recovered motion corrections; their nearly horizontal tracks indicate successful range-migration compensation. Colours show power relative to background, measured before the search for (a,b) and after filtering for (c,d). Panels (e,f) show recovered power versus trial radial velocity, normalized to its maximum. Broad peaks indicate limited velocity precision. Dashed lines mark catalogue velocities, added for comparison after the search. Time is relative to each blind-fit epoch.}
\label{fig:blind}
\end{figure}
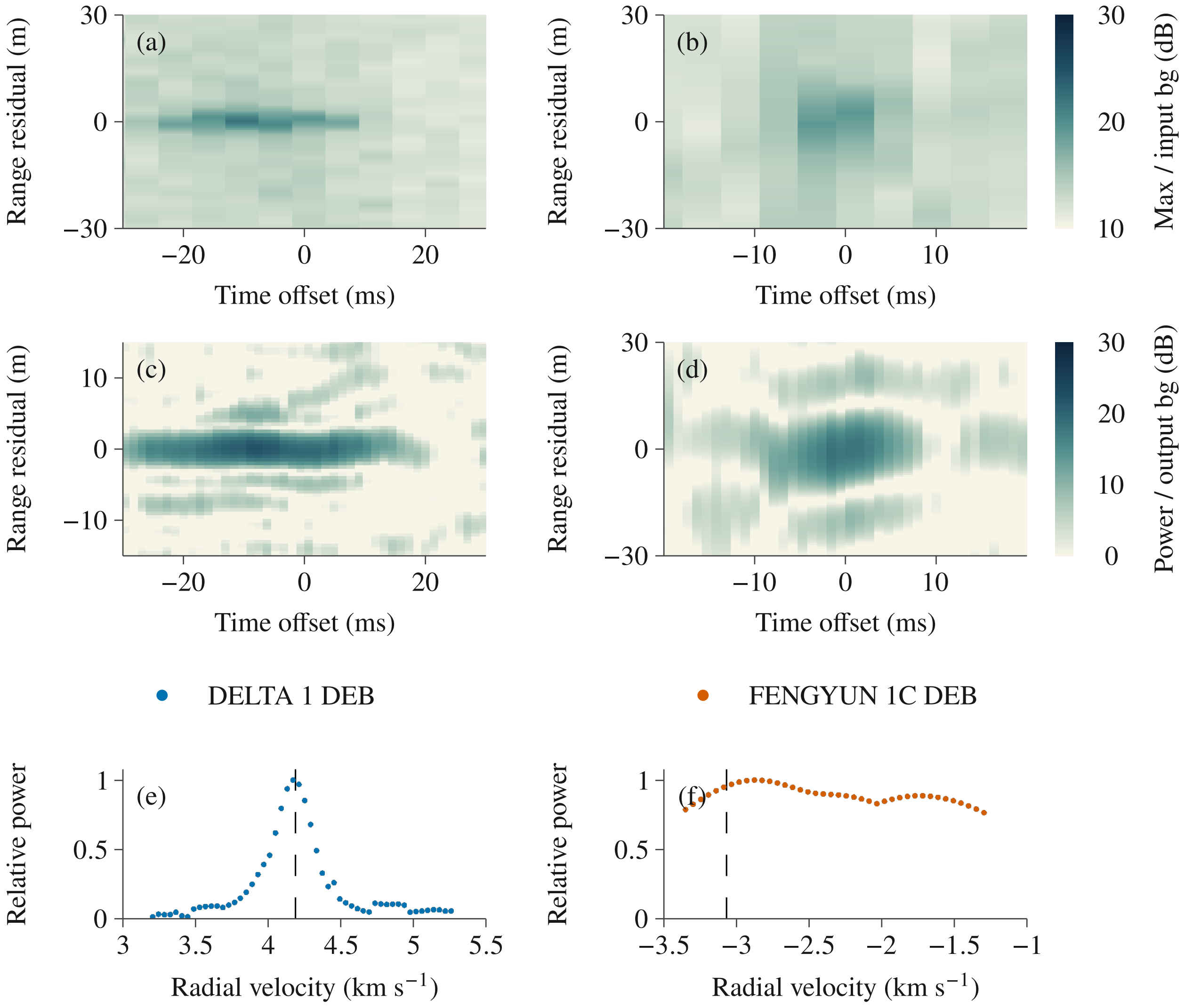

The blind search tests whether the debris echoes can be found without knowing their trajectories beforehand. It recovers both DELTA 1 DEB and FENGYUN 1C DEB (Fig.~\ref{fig:blind}). We search the recorded range intervals over 1.655 s and 0.553 s, respectively, including times outside the object transits (Table~\ref{tab:blind-domain}). Catalogue comparison and manual inspection then identify the debris echoes among the candidates.
Figure~\ref{fig:blind} follows each detection from the search to the recovered echo. Panels (a,b) retain the strongest response across the tested motions and Doppler frequencies. This also selects stronger background fluctuations, making the search maps appear noisier. Panels (c,d) then apply the selected motion correction: each echo remains at nearly constant range, with peak SNRs of 22.1 dB for DELTA 1 DEB and 18.3 dB for FENGYUN 1C DEB.
For DELTA 1 DEB, the blind search recovers essentially the same echo power as catalogue-assisted processing. The recovered radial velocity is approximately consistent with the catalogue prediction. However, the broad responses in panels (e,f) show that a range of trial velocities recovers similar power, limiting the velocity precision.
\begin{table}[tbp]
\centering\small
\caption{Blind-search coverage and detection results for the two debris encounters in Fig.~\ref{fig:blind}. ``Analyzed'' gives the observation duration, ``CPIs'' the number of overlapping coherent processing intervals, and ``Gate'' the recorded range interval searched. Each search recovers one known debris object. SNR is the peak echo strength after correcting for the recovered motion. False-positive rates count ground-scatter peaks above the 6 dB search threshold per observation second, counting overlapping time intervals once. A recurring ground response can contribute multiple peaks.}
\label{tab:blind-domain}
\setlength{\tabcolsep}{4pt}
\begin{tabular}{@{}lrrrrrrr@{}}\toprule
Acquisition & UTC start & Analyzed (s) & CPIs & Gate (km) & Known & SNR (dB) & False peaks/s\\\midrule
DELTA 1 DEB & 02:07:02.091 & 1.655 & 299 & 26.24--91.63 & 1 & 22.1 & 408\\
\shortstack[l]{FENGYUN 1C\\DEB} & 19:47:54.984 & 0.553 & 131 & 13.18--66.89 & 1 & 18.3 & 0\\\bottomrule
\end{tabular}
\end{table}
Ground scatter is the main obstacle to automatic detection. At the 6 dB threshold, the searches produce 676 false-positive ground-scatter peaks across 2.208 s of observations, approximately 306 peaks/s. All occur in the DELTA 1 DEB excerpt, giving 408 peaks/s there. A persistent ground response near 88.60 km nominal gate range crosses threshold in 247 processing windows, so many peaks are repeated detections of the same feature. The FENGYUN 1C DEB search produces no false-positive peaks in the analyzed interval due to more favorable ground scatter during the beam crossing. Rates count overlapping observation intervals once.
\FloatBarrier
\subsection{Orbit refinement of DELTA 1 DEB}
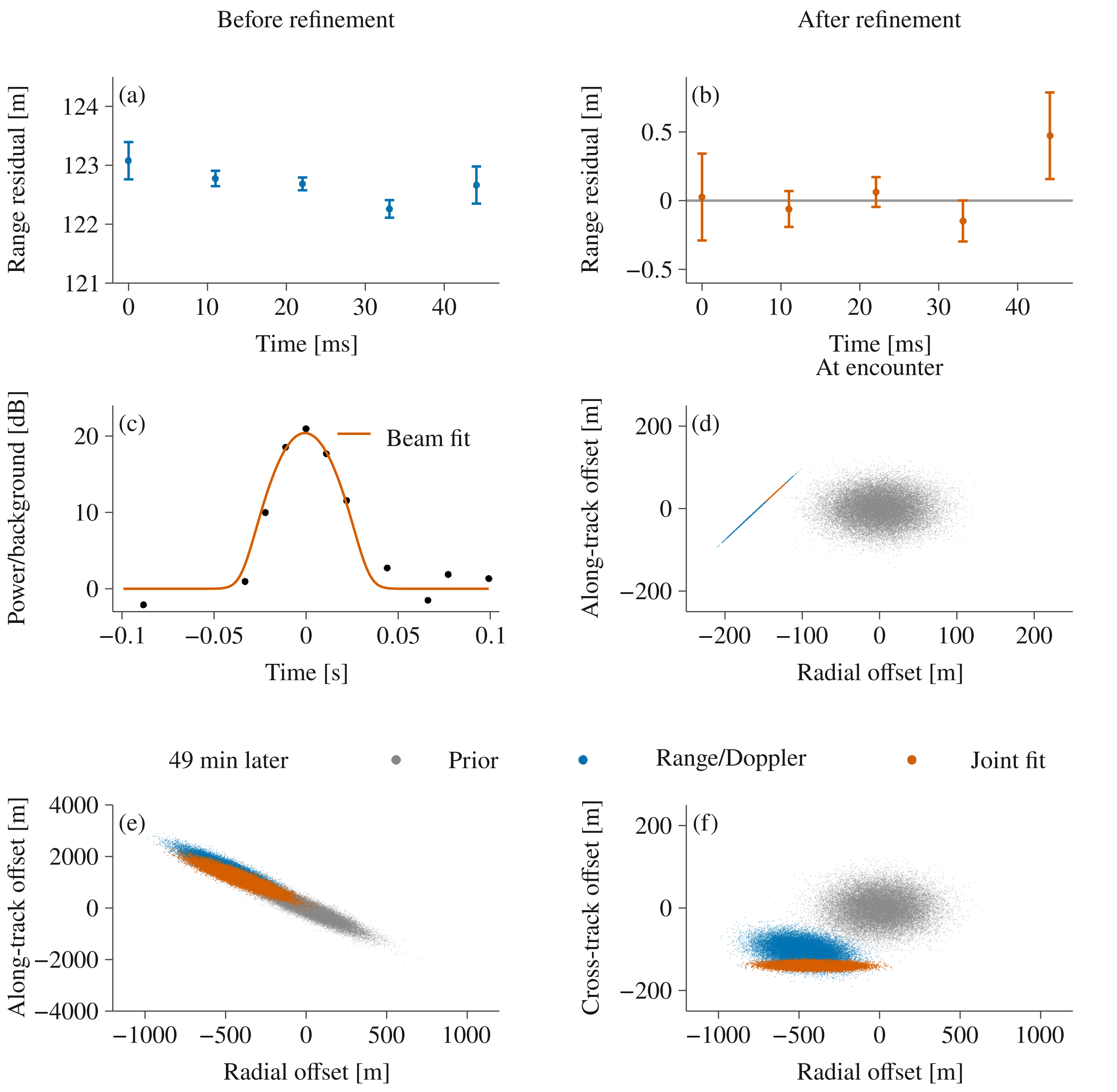
\begin{figure}[tbp]
\centering\input{paper/figures/debris-orbit-refinement.include.tex}
\caption{DELTA 1 DEB orbit refinement assuming known TOPS pointing. (a,b) Range residuals before (left) and after (right) the joint orbit refinement, using the same five measurement windows and separate vertical scales. Error bars show waveform-derived $1\sigma$ range uncertainties; time is measured from the first window. (c) Echo power and beam-pattern fit, relative to the echo peak. (d) Sampled positions at the encounter. (e,f) Positions half an orbit later in radial--along-track and radial--cross-track coordinates relative to the catalogue position. Grey shows the prior, blue the range--Doppler update and orange the joint update including the beam passage.}
\label{fig:debris-orbit}
\end{figure}

Range and Doppler constrain the object's distance and motion along the radar line of sight. The beam passage adds angular information: when the echo brightens and fades helps locate the object within the narrow antenna beam. Figure~\ref{fig:debris-orbit} compares orbit updates using range and Doppler alone with a joint fit that also uses this echo envelope, assuming the TOPS pointing history is known.
At the encounter, the joint fit reduces the three-dimensional RMS position spread from 59.4 m in the adopted prior to 8.5 m. The benefit also persists when the sampled orbits are propagated forward. Half an orbit later, approximately 49 min, range and Doppler alone reduce the prior spread from 500 m to 406 m; including the beam passage reduces it to 354 m, a 29\% reduction. The joint fit reduces the range residual RMS from 123 m to 0.22 m and the Doppler residual RMS from 701 Hz to 5.6 Hz, equivalent to radial-velocity residuals of 19.4 and 0.16 m/s. The post-fit range residual is approximately one twelfth of the recorded 2.746 m sample spacing. Fractional-delay interpolation and coherent waveform fitting locate the echo between samples, with residuals comparable to the estimated per-window range uncertainties of 0.11--0.32 m.
The updated future position is displaced by 1.22 km from the catalogue prediction. This displacement is the correction to the predicted position; the quoted spreads describe uncertainty around each distribution's own mean. Under the adopted pointing, noise and prior assumptions, this example shows how a calibrated beam passage can improve the future position estimate.
\FloatBarrier
\subsection{Ideal sensitivity of orbital radars}
\begin{figure}[htbp]\centering
\input{figures/receiver-sensitivity.include.tex}
\caption{Receiver-noise-equivalent conducting-sphere diameter versus range for the radars in Table~\ref{tab:radars}. Curves assume unit SNR, central crossings at 10 km/s and coherent integration limited by beam dwell or 10 ms, with the recording, gain and loss assumptions in Methods. Discontinuities arise from sphere resonances.}
\label{fig:noise-diameter}
\end{figure}
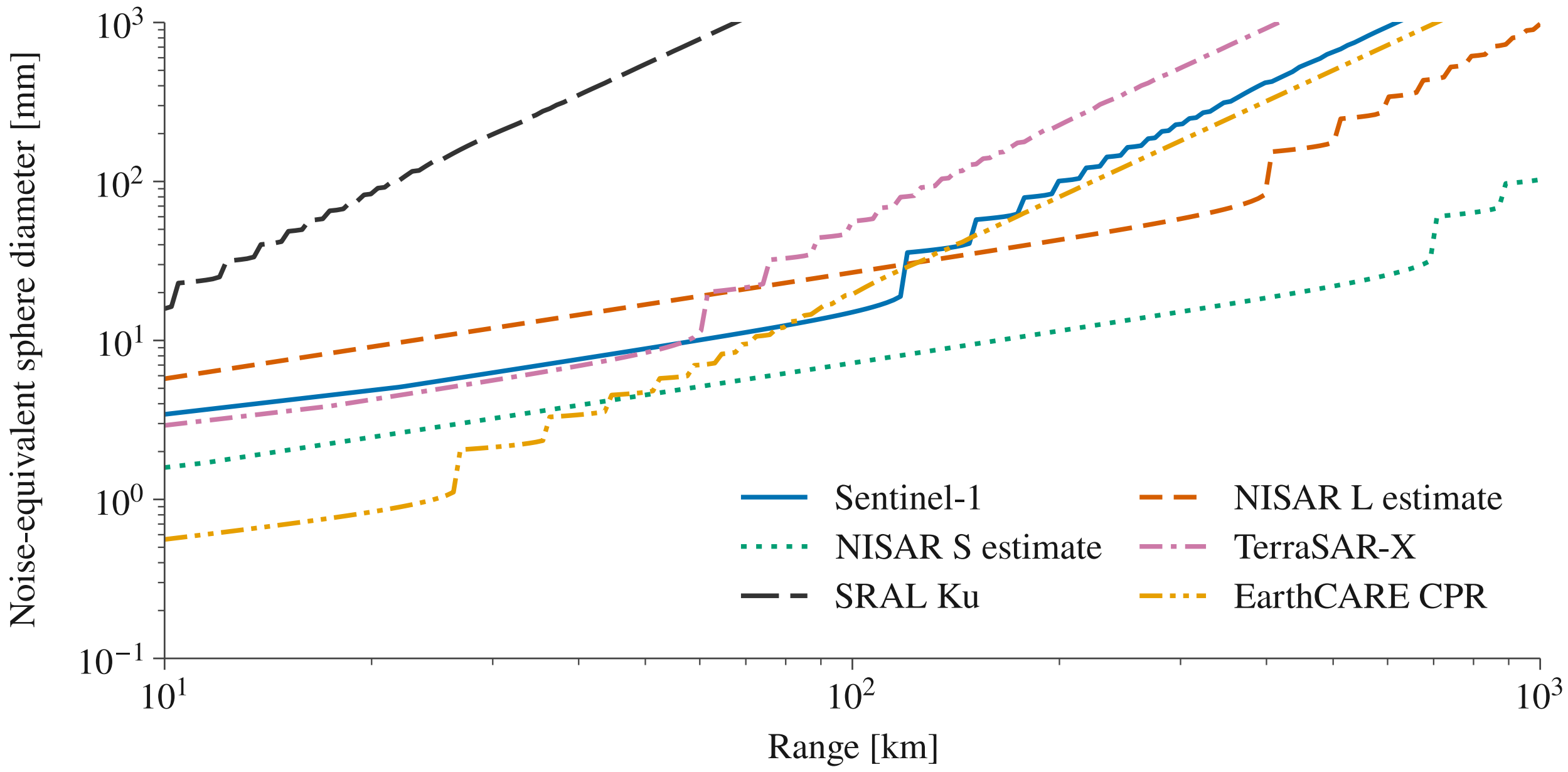

Figure~\ref{fig:noise-diameter} shows how the smallest sphere reaching unit SNR changes with distance under the receiver-noise assumptions of Section~\ref{sec:receiver-sensitivity}. Lower curves indicate sensitivity to smaller objects. At 10 km, the calculated diameters are 3.43 mm for Sentinel-1 and 0.56 mm for EarthCARE. Table~\ref{tab:radars} gives the values at 10 and 50 km for all instruments. The jumps in the curves arise from resonances in the radar scattering from a conducting sphere.
These estimates account for the short time an object remains in the beam. At 10 km, Sentinel-1 provides 4.54 ms of coherent integration, whereas EarthCARE's narrower beam provides 1.66 ms. Their transmit duty factors reduce the accumulated transmitted pulse durations to 0.408 ms and 0.0356 ms, respectively. EarthCARE nevertheless reaches a smaller noise-equivalent diameter, illustrating how instrument sensitivity can outweigh a shorter beam crossing. Sentinel-1 reaches the assumed 10 ms integration cap near 22 km; beyond this range, a longer crossing provides no additional integration in this comparison.
\FloatBarrier
\section{Discussion}
\subsection{Space-object sensitivity in Earth-observation measurements}
Figure~\ref{fig:delta-envelope} shows why echo strength alone is an incomplete guide to object size. The two polarization channels follow the same beam passage but can differ substantially in amplitude, particularly for the two debris fragments and SL-14 R/B. An estimated RCS below the historical NORAD class interval can reflect the object's polarization and viewing geometry, motion-compensation losses, or uncertainty in antenna gain and background power.
Our RCS estimates generally fall below the historical NORAD class ranges, particularly for weak echoes (Fig.~\ref{fig:human-rcs}). Ground scatter could contribute by raising the background above the thermal-noise level assumed in the RCS calculation. A background ten times the receiver-noise power would require a 10 dB upward RCS correction, with other assumptions unchanged. This multiplicative correction applies equally to weak and strong echoes. Overestimating the subtracted background, however, would disproportionately suppress weak echoes and could explain part of their larger discrepancy.
Imperfect motion compensation can also reduce echo power, especially in nearby encounters with rapid range migration. Independent background, receiver-noise and antenna-gain calibration would help distinguish these effects from variations caused by viewing geometry and polarization.
\subsection{Blind searches, clutter and encounter dynamics}
The blind search recovers both debris echoes, but ground scatter complicates automatic identification. The DELTA excerpt produces many clutter peaks, whereas the FENGYUN 1C DEB excerpt produces none at the selected threshold (Table~\ref{tab:blind-domain}). Acquisitions and the threshold were selected using known encounters; catalogue comparison and manual inspection provide the final identifications.
Ground and orbital echoes can overlap in sampled Doppler. TOPS steering changes ground-return Doppler, while pulse sampling folds frequencies into the recorded interval~\cite{deramp}. Distinguishing the echoes requires following their range and Doppler through time and checking which transmitted pulses could have produced the recorded returns. Adaptive filtering could suppress ground scatter while preserving short orbital echoes.
Distant large satellites can resemble nearby small debris when detected through sidelobes or assigned the wrong pulse order. Most catalogue-correlated detections in the 2018 EISCAT Troms\o{} and Svalbard beam-park experiment arrived through sidelobes~\cite{vierinenBeampark2018}. Assuming boresight gain underestimates their RCS. Catalogue correlations, antenna-pattern models and consistent range--Doppler histories are therefore important for interpreting object sizes and debris counts.
Nearby encounters also require accurate motion compensation: rapid line-of-sight rotation can produce large range acceleration even when instantaneous Doppler is small (Eq.~\ref{eq:accel}). Short coherent windows limit this evolution; longer integrations must track the changing range and Doppler, including higher-order motion where necessary. Orbital population models could help restrict the trial motions and pulse-order alternatives searched.
\subsection{Orbit refinement and catalogue updates}
Routine Sentinel-1 acquisitions could provide additional tracking measurements without changing the Earth-observation schedule. Published delivery targets include Level-0 availability within 3 h for designated near-real-time areas and 24 h globally~\cite{s1Timeliness}. The additional work would be processing recordings already acquired and downlinked.
A pilot service could predict catalogue-object passages through recorded beams and deliver fitted range and Doppler measurements with uncertainties. Figure~\ref{fig:debris-orbit} shows how the beam-passage timing could add an angular constraint, provided the antenna pattern and TOPS pointing are accurately calibrated. Combining these measurements with independent tracking would establish their contribution to catalogue accuracy.
Conjunction warnings involving high-value spacecraft offer a practical priority. Available and planned acquisitions could be screened for passages of either object before the collision-avoidance decision is due. Radar measurements combined with existing tracking and operator ephemerides could then refine the predicted miss distance and collision risk~\cite{nasaConjunctionHandbook}. Their operational value would depend on viewing geometry, delivery and processing time, measurement calibration and resolution of Doppler ambiguities.
\subsection{Potential instruments and statistical debris surveys}
The sensitivity comparison identifies W-band cloud radars as promising instruments for millimetre-scale debris measurements (Fig.~\ref{fig:noise-diameter}). Reaching the calculated sensitivity requires receive windows covering nearby objects and recording coherent pulse voltages. For EarthCARE, this would mean adapting the recording settings used for atmospheric observations near the surface~\cite{earthcareSpec}.
Observations directed away from Earth would reduce ground scatter and bring sensitivity closer to the receiver-noise limit. Existing or dedicated spaceborne radars could then measure debris fluxes through statistical surveys, analogous to ground-based beam parks~\cite{ordemProcess,vierinenBeampark2018}. Converting detection counts into flux requires the observing duration, volume sampled and probability of detecting objects of different sizes and trajectories.
The wide elevation beam and narrow azimuth beam of SAR instruments could also help estimate object size. For passages near the elevation-beam centre, fitting the azimuth crossing would constrain the changing antenna gain and improve the RCS estimate. Interferometric measurements could further constrain the object's position within the beam.
\FloatBarrier
\section{Conclusions}
Sentinel-1 radar voltages acquired for Earth observation contain measurable echoes from satellites, rocket bodies and debris fragments. Catalogue-assisted processing identifies 76 objects at ranges of 41.7--492.9 km, and a blind search recovers both tested debris transits. For DELTA 1 DEB, blind refinement recovers essentially the same coherent power as catalogue-assisted processing. An idealized orbit update also shows how range, Doppler and calibrated beam-passage timing can improve the predicted position after a short encounter.
Ground scatter is the main limitation for blind detection in these Earth-looking recordings. Accurate motion compensation, clutter suppression that preserves orbital echoes, and independent gain and noise calibration would improve detection and RCS estimation. Resolving range and Doppler ambiguities would further support reliable object association and orbit refinement.
The receiver-noise sensitivity comparison indicates potential for millimetre-to-centimetre debris surveys, particularly with W-band radars. Receive windows adapted to nearby objects, coherent voltage recording and observations directed away from Earth could help realize this potential. With measured detection efficiency and observing exposure, such surveys could constrain debris fluxes as ground-based beam parks do today. Processing routine Earth-observation recordings could also supply catalogue updates, while targeted radar observations could provide additional tracking for objects involved in conjunction warnings.
\section*{Data and software availability}
The source measurements are publicly available Copernicus Sentinel-1 Level-0 products. Contains modified Copernicus Sentinel data [2026]. Orbital elements are obtained from Space-Track under its access conditions. Analysis software and reduced products are maintained in a private author repository until manuscript publication.
\section*{Declarations}
\paragraph{Author contributions.} Juha Vierinen conceived the study, developed the methodology and software, performed the analysis, produced the figures and wrote the original manuscript. All authors contributed to the interpretation of the results, critically reviewed the manuscript and approved the final version.
\paragraph{Funding:} Part of the work was funded by the Research Council of Norway, projects \#326609 and \#335832, as well as the UiT grant \#2531957.
\paragraph{Declaration of competing interests.} The authors declare no competing financial interests or personal relationships that could have influenced this work.
\section*{Declaration of generative AI and AI-assisted technologies in the manuscript preparation process}
During preparation of this work, the authors used GPT-6.1 to assist with drafting and code development. The authors reviewed and edited the outputs and take responsibility for the scientific interpretation and final manuscript.
\bibliographystyle{elsarticle-num}
\bibliography{paper/references}
\end{document}

%% file: paper/figures/sentinel-debris.include.tex
\begingroup
\normalsize
\dimen0=\linewidth
\advance\dimen0 by -174.000000mm
\ifdim\dimen0<0pt \dimen0=-\dimen0\fi
\ifdim\dimen0>0.1mm
  \PackageError{articlefigure}{Article linewidth changed}{Regenerate this figure using the new ArticleLayout.}
\fi
\dimen0=\csname f@size\endcsname pt
\advance\dimen0 by -10.950000pt
\ifdim\dimen0<0pt \dimen0=-\dimen0\fi
\ifdim\dimen0>0.01pt
  \PackageError{articlefigure}{Article body font changed}{Regenerate this figure using the new ArticleLayout.}
\fi
\includegraphics[width=174.000000mm]{\detokenize{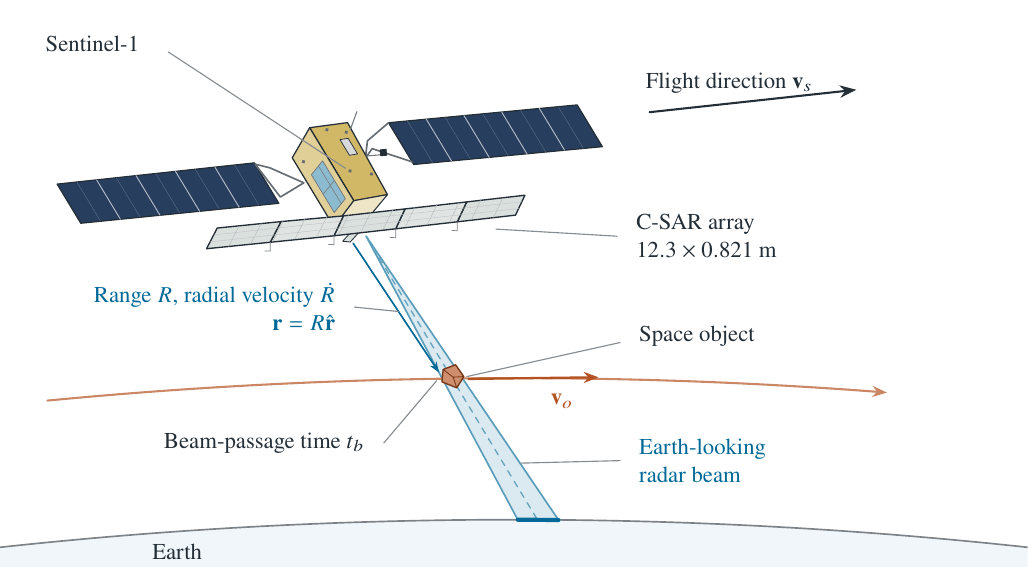}}
\endgroup

%% file: paper/data/human-validated-rcs-counts.tex
\def\HumanRcsObjectCount{76}
\def\HumanRcsMeasurementCount{147}

%% file: paper/figures/assisted-comparison.include.tex
\begingroup
\normalsize
\dimen0=\linewidth
\advance\dimen0 by -174.000000mm
\ifdim\dimen0<0pt \dimen0=-\dimen0\fi
\ifdim\dimen0>0.1mm
  \PackageError{rustplotlib}{Article linewidth changed}{Regenerate this figure using the new ArticleLayout.}
\fi
\dimen0=\csname f@size\endcsname pt
\advance\dimen0 by -10.950000pt
\ifdim\dimen0<0pt \dimen0=-\dimen0\fi
\ifdim\dimen0>0.01pt
  \PackageError{rustplotlib}{Article body font changed}{Regenerate this figure using the new ArticleLayout.}
\fi
\includegraphics[width=164.559997mm]{\detokenize{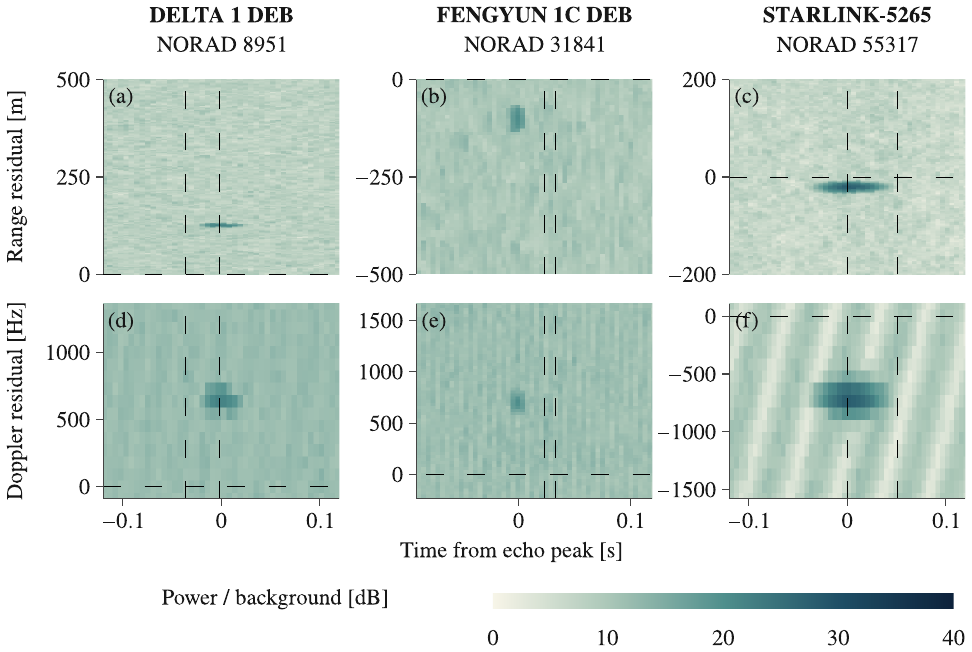}}
\endgroup

%% file: paper/figures/beam-examples.include.tex
\begingroup
\normalsize
\dimen0=\linewidth
\advance\dimen0 by -174.000000mm
\ifdim\dimen0<0pt \dimen0=-\dimen0\fi
\ifdim\dimen0>0.1mm
  \PackageError{rustplotlib}{Article linewidth changed}{Regenerate this figure using the new ArticleLayout.}
\fi
\dimen0=\csname f@size\endcsname pt
\advance\dimen0 by -10.950000pt
\ifdim\dimen0<0pt \dimen0=-\dimen0\fi
\ifdim\dimen0>0.01pt
  \PackageError{rustplotlib}{Article body font changed}{Regenerate this figure using the new ArticleLayout.}
\fi
\includegraphics[width=170.486671mm]{\detokenize{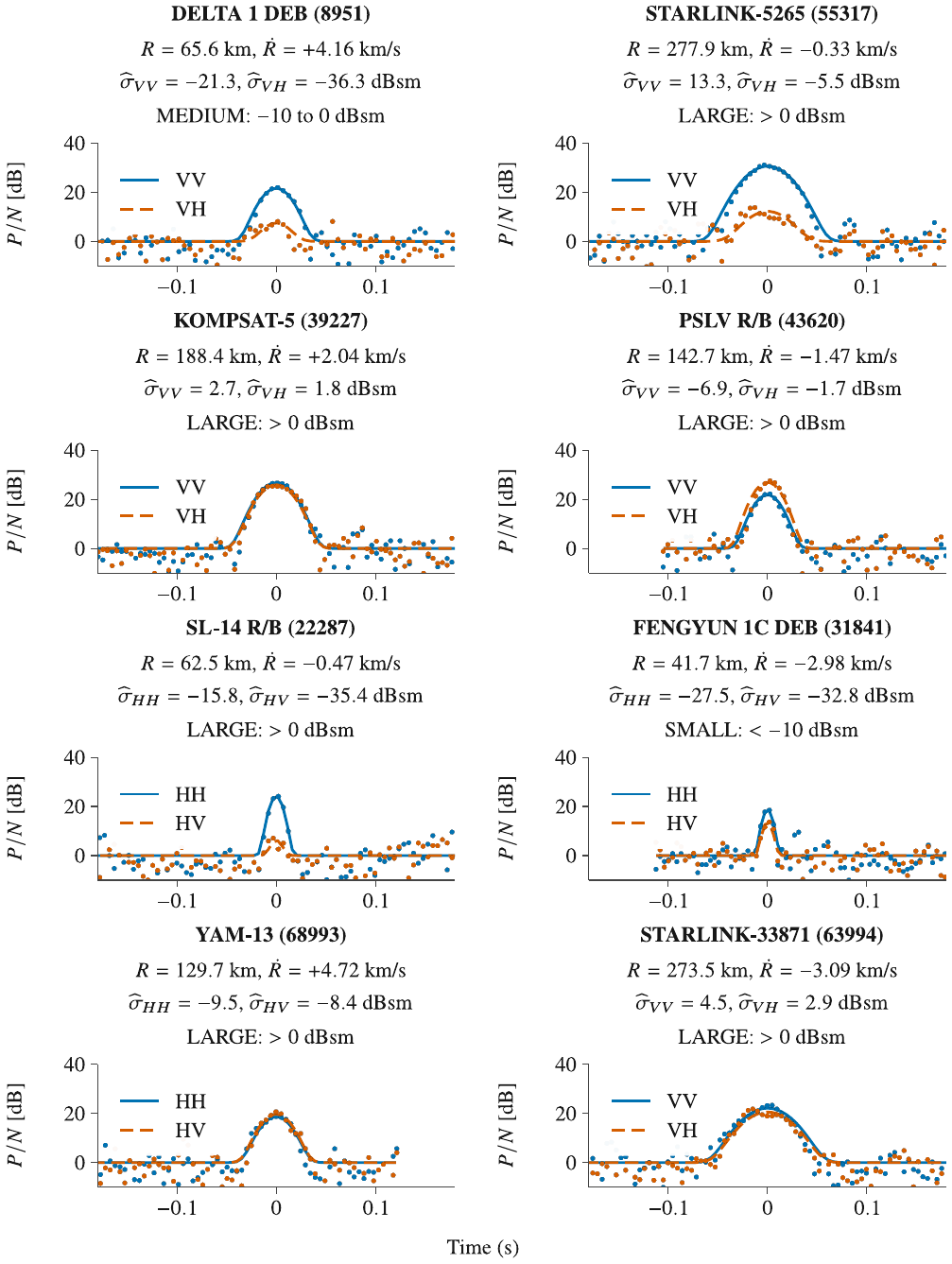}}
\endgroup

%% file: paper/figures/human-validated-rcs.include.tex
\begingroup
\normalsize
\dimen0=\linewidth
\advance\dimen0 by -174.000000mm
\ifdim\dimen0<0pt \dimen0=-\dimen0\fi
\ifdim\dimen0>0.1mm
  \PackageError{rustplotlib}{Article linewidth changed}{Regenerate this figure using the new ArticleLayout.}
\fi
\dimen0=\csname f@size\endcsname pt
\advance\dimen0 by -10.950000pt
\ifdim\dimen0<0pt \dimen0=-\dimen0\fi
\ifdim\dimen0>0.01pt
  \PackageError{rustplotlib}{Article body font changed}{Regenerate this figure using the new ArticleLayout.}
\fi
\includegraphics[width=163.374669mm]{\detokenize{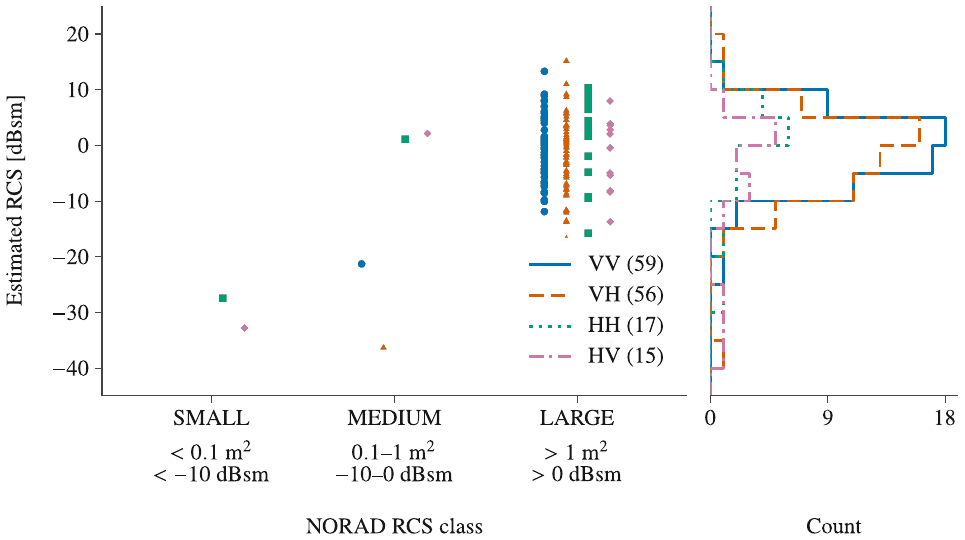}}
\endgroup

%% file: paper/figures/blind-comparison.include.tex
\begingroup
\normalsize
\dimen0=\linewidth
\advance\dimen0 by -174.000000mm
\ifdim\dimen0<0pt \dimen0=-\dimen0\fi
\ifdim\dimen0>0.1mm
  \PackageError{rustplotlib}{Article linewidth changed}{Regenerate this figure using the new ArticleLayout.}
\fi
\dimen0=\csname f@size\endcsname pt
\advance\dimen0 by -10.950000pt
\ifdim\dimen0<0pt \dimen0=-\dimen0\fi
\ifdim\dimen0>0.01pt
  \PackageError{rustplotlib}{Article body font changed}{Regenerate this figure using the new ArticleLayout.}
\fi
\includegraphics[width=166.761333mm]{\detokenize{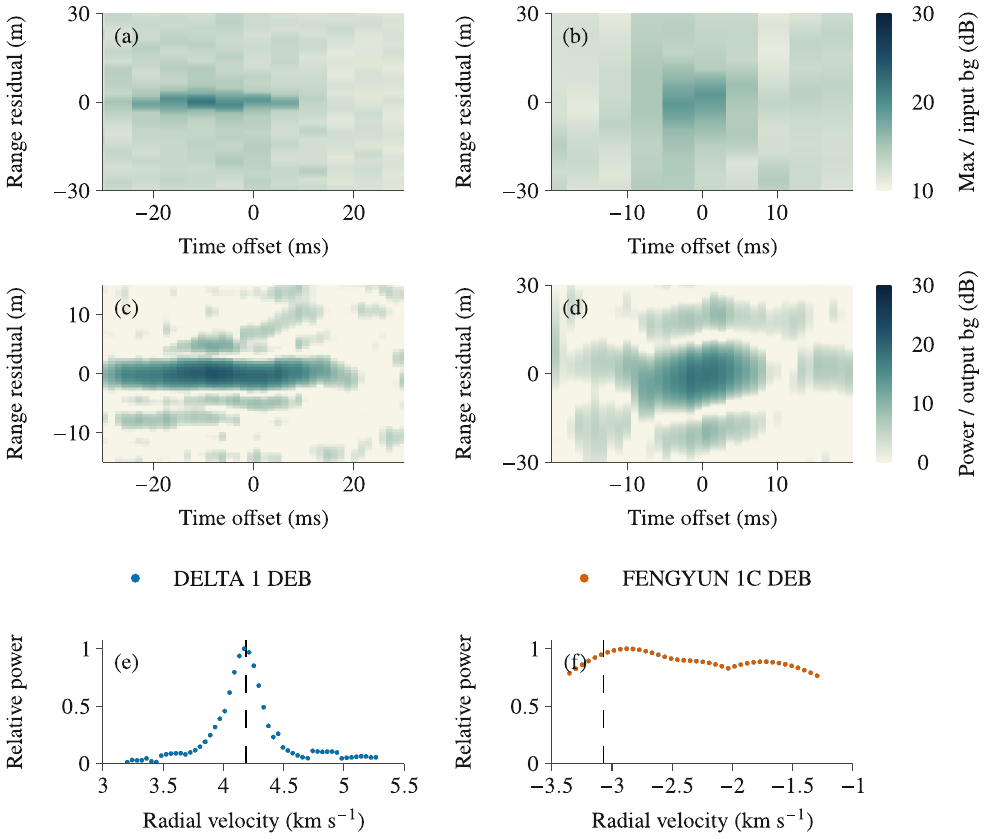}}
\endgroup

%% file: paper/figures/debris-orbit-refinement.include.tex
\begingroup
\normalsize
\dimen0=\linewidth
\advance\dimen0 by -174.000000mm
\ifdim\dimen0<0pt \dimen0=-\dimen0\fi
\ifdim\dimen0>0.1mm
  \PackageError{rustplotlib}{Article linewidth changed}{Regenerate this figure using the new ArticleLayout.}
\fi
\dimen0=\csname f@size\endcsname pt
\advance\dimen0 by -10.950000pt
\ifdim\dimen0<0pt \dimen0=-\dimen0\fi
\ifdim\dimen0>0.01pt
  \PackageError{rustplotlib}{Article body font changed}{Regenerate this figure using the new ArticleLayout.}
\fi
\includegraphics[width=167.438665mm]{\detokenize{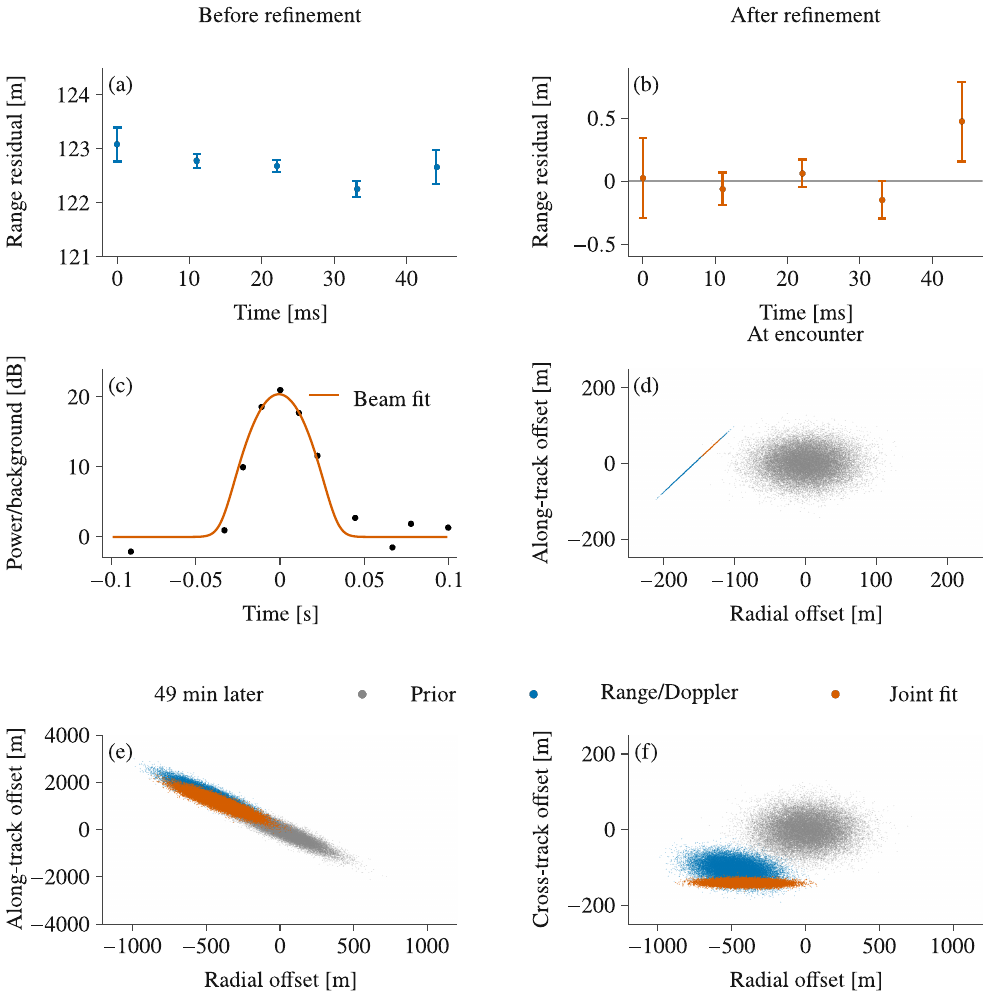}}
\endgroup

%% file: figures/receiver-sensitivity.include.tex
\begingroup
\normalsize
\dimen0=\linewidth
\advance\dimen0 by -174.000000mm
\ifdim\dimen0<0pt \dimen0=-\dimen0\fi
\ifdim\dimen0>0.1mm
  \PackageError{rustplotlib}{Article linewidth changed}{Regenerate this figure using the new ArticleLayout.}
\fi
\dimen0=\csname f@size\endcsname pt
\advance\dimen0 by -10.950000pt
\ifdim\dimen0<0pt \dimen0=-\dimen0\fi
\ifdim\dimen0>0.01pt
  \PackageError{rustplotlib}{Article body font changed}{Regenerate this figure using the new ArticleLayout.}
\fi
\includegraphics[width=165.745326mm]{\detokenize{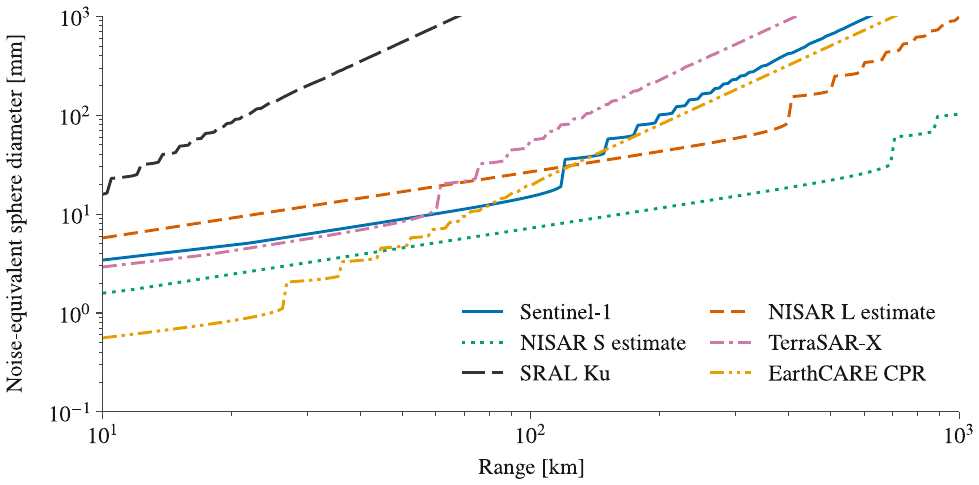}}
\endgroup